\documentclass[conference]{IEEEtran}
\usepackage{amsmath}
\usepackage[OT1]{fontenc} 

\usepackage{cite}
\usepackage{amsmath,amssymb,amsfonts}
\usepackage{algorithmic}
\usepackage{graphicx}
\usepackage{textcomp}
\usepackage{xcolor}
\usepackage{todonotes}
\def\BibTeX{{\rm B\kern-.05em{\sc i\kern-.025em b}\kern-.08em
    T\kern-.1667em\lower.7ex\hbox{E}\kern-.125emX}}
\usepackage{amssymb}
\usepackage[linesnumbered,lined,boxed,commentsnumbered]{algorithm2e}
\usepackage{amsthm}
\theoremstyle{remark}
\newtheorem{definition}{Definition}[section]

\usepackage{subcaption}
\usepackage{tcolorbox}

\newtheoremstyle{myaxiom}
  {0.3\baselineskip plus 0.2ex minus 0.1ex}
  {0.3\baselineskip plus 0.2ex minus 0.1ex}
  {\itshape}
  {}
  {\bfseries}
  {}
  {1em}
  {\thmname{#1}\normalfont\thmnote{ (#3)}}%

\theoremstyle{myaxiom}
\newtheorem*{myaxiom}{\myaxiomname}
\newcommand{\myaxiomname}{}

\newenvironment{axiom}[1]
 {\renewcommand\myaxiomname{#1}\myaxiom}
 {\endmyaxiom}

\begin{document}

\title{Towards a Foundational Ontology for Identifying and Resolving Contradictions in Dialogue-based Human-Robot Interactions
}

\author{\IEEEauthorblockN{Maitreyee Tewari}
\IEEEauthorblockA{\textit{Nord A1, Sweden} \\
\textit{Alternative Intelligence LLP,}\\
Bologna, Italy \\
ORCID: 0000-0002-3036-6519}
\and
\IEEEauthorblockN{Michele Persiani}
\IEEEauthorblockA{\textit{Department of Informatics} \\
\textit{University of Bologna}\\
Bologna, Italy \\
ORCID: 0000-0001-5993-3292}
}

\maketitle

\begin{abstract}
Existing Human-Robot Interaction (HRI) literature has focused on identifying and structuring errors, failures, conflicts, and knowledge issues (called in this work as contradictions) in domain-specific dialogue-based interactions. However, there is still lack of a formal computational framework to represent and define these contradictions, interoperable and usable across HRI and human-agent interaction (HAI) domains. Thus, this research project aims to capture, represent, and evaluate the notion of (1) dialogue-based collaborative interaction and (2) related contradictions in a foundational ontology. METHONTOLOGY, a systematic approach to build domain-independent ontologies was applied. In the conceptualisation stage of the presented ontology, concepts and models from Activity Theory were used. Preliminary results presented in this short article are: (i) Natural language definitions of dialogues and related contradictions in HRI, (ii) Set Theoretic definitions of dialogues and contradictions, and (iii) First Order Logic (FoL) formulation of the contradiction concepts and three novel principles guiding dialogue-based interactions between humans and robots. In summary, we report on ongoing work to develop a foundational ontology based on Activity Theory called Activity Theory-based foundational ontology (ATFOt) to capture and represent the notion of contradictions in HRI.
\end{abstract}

\begin{IEEEkeywords}
Contradictions,Activity Theory, Engeström's Activity System Model, Human-Robot Interaction, Knowledge Representation, Foundational Ontologies, First Order Logic, Set Theory.
\end{IEEEkeywords}

\section{Introduction}\label{sec:intro}
Large Language Models (LLM) have been shown to generate coherent and contextual text similar to human level language generation capabilities. LLMs integrated in virtual assistants (VAs) such as Alexa, Siri, and ChatGPTs have surpassed human expectations in query-based interactions\cite{MAHMOOD2025103406}. This success of LLMs with VAs prompted the sudden and rapid integration of LLMs in dialogue-based HRI\cite{WenEtAl_2024,Ilda_etAl_24,MAHMOOD2025103406}. However, this integration of LLMs in HRI merely underscored the already existing limitations such as lack of understanding about the semantics, intent, and norms, lack of contextual knowledge, strategies to manage errors and uncertainties  (HRI)\cite{Ilda_etAl_24,CaoEtAL_2025}. Furthermore, researcher's\cite{ToneyWails_Singh_2024} found that LLM technology's limitation of \textit{confabulation,} (i.e., to generate seemingly coherent yet false and manufactured information\cite{SmithEtAl_2023}) coupled with inconsistent and lack of uncertainty expressions makes them incompetent for ethical HRI. These findings adds to the already existing complex problems of unpredictability in dialogue-unfolding that generally causes communication failures, speech errors, knowledge issues, and conflicts in dialogue-based HRI\cite{TianLeimin2021,tewari2023breakdown,CaoEtAL_2025,KoxEtAl_2025,NogueiraEtAl2026dont}.  

To address failures, knowledge issues, and conflicts, researchers have proposed taxonomies representing HRI failures related to capabilities of robots\cite{NogueiraEtAl2026dont}, social errors\cite{TianLeimin2021}, and conflicts and knowledge issues\cite{TewaLindFrontiers2022}. Methods to identify and manage failures through speech and gestures\cite{WachowiakEtAl2024}, interaction patterns\cite{MAHMOOD2025103406}; conflict of intentions using Answer Set Programming\cite{GuerreroEtAl_2023} have been proposed. Given this prior work, however, there is still a lack of structured and granular knowledge representation of communication errors and failures, conflicts, and knowledge issues arising in dialogue-based HRI.

Therefore, the objective of this research is to formally represent the knowledge of dialogue-based interactions (HRI and HAI) and related communication errors, conflicts, and knowledge issues with a foundational ontology. Following the naming convention from Activity Theory, that describes human activity as purposeful, evolving and transforming the actor and the subjective world\cite{leontiev1981,Vygotsky1982,KaptelininNardi2012ATHCI},) this work refers to errors, knowledge issues, and conflicts as \textit{contradictions}\cite{engestrom1999ActivityCh2,Engestrom_Sannino_CHAT_2018,Engestrom_Miettinen_Punamaki_1999}. In the rest of the article, knowledge issues and conflicts will be called collectively as contradictions.    

The work presented in this article, partially fulfilling the research objective, explores the research question: 

\textit{How can we create a general representation of contradictions in dialogue-based HRI within a foundational ontology?}

To address the research question, activity analysis using the Activity System Model and the concept of \textit{contradictions} from activity theory was conducted on HRI \cite{tewari2023breakdown} and HAI (digital and virtual agent)\cite{kaelin2024developing,LindgrenEtAl2024_VOT} scenarios. These scenarios embedded varying types of contradictions. The analysis informed the development of a \textit{foundational} ontology encoding a hierarchy of contradictions representing errors, conflicts, and knowledge issues for dialogue-based HRI and HAI. The hierarchy categorises these conflicts and knowledge issues at four levels of abstraction (similar to the four levels of contradictions in activity theory).   

    
The contribution of this article is a \textit{novel formal knowledge representation of contradictions} and \textit{three novel principles} guiding and transforming the dialogue-based interactions and the involved participants. 
    
The rest of the article is structured as follows. Section \ref{sec:method} briefly presents the methodology being applied to develop the ontology. The formal specification of contradictions using Set Theory and First Order Logic (FoL) and relevant definitions from the ontology are presented in Section \ref{sec:formal_rep}. Section \ref{sec:future_work} concludes the article, informing the status and research contributions of the work.   


\section{Methodology}\label{sec:method}
To develop the ontology, METHONTOLOGY\cite{methontology97} provides six activities that need to be performed: (1) specification, (2) knowledge acquisition, (3) conceptualisation, (4) integration, (5) implementation, and (6) evaluation. 

The specification activity defines the purpose and use-case, the level of formality, scope, characteristics, and granularity of the ontology. The outcome is a specification document in natural language. 

Conceptualisation structures the domain knowledge in a conceptual model describing the problem and the solution in the specification document. 

During integration, concepts of the developed conceptual model are mapped to existing meta ontologies. The outcome is a document associating the terms in the referenced meta ontologies to model's concepts. 

The development activity commences by formally representing the concepts using logic-based methods. Then a language and an appropriate environment to develop the ontology is selected. We are using Set Theory and First Order Logic to formally define and represent the contradictions concepts. Protege is being used to implement and evaluate the ontology. 

In the evaluation activity, the ontology is inspected by verifying its correctness (verification process) and validating that the ontology represents the intended system knowledge \cite{methontology97}. 

\begin{figure}
    \centering
    \includegraphics[width = 0.5\textwidth]{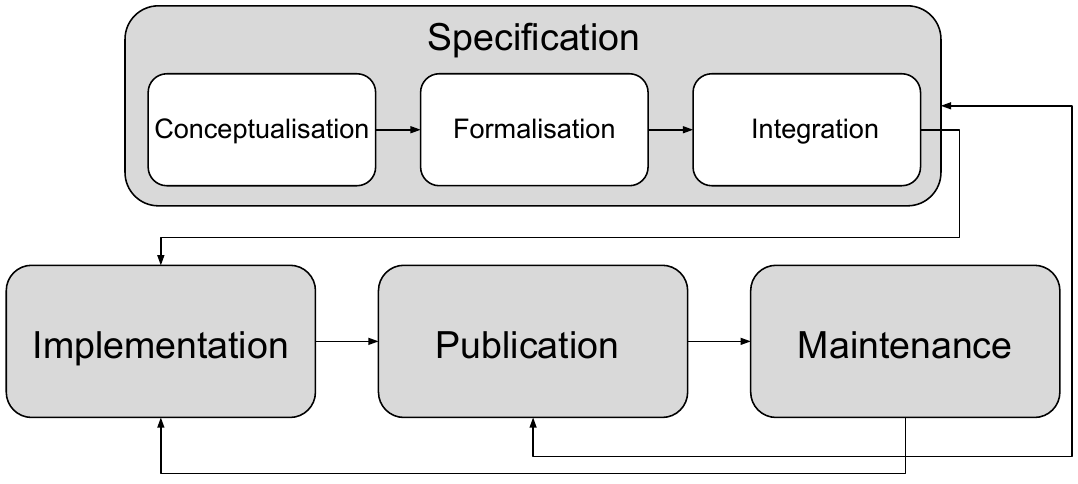}
    \caption{Central phases of METHONTOLOGY illustrated in white boxes. The other additional phases of knowledge acquisition, implementation, and documentation are illustrated in grey with the darker gradient implying higher activity in the corresponding phases. The figure is an adaptation from \cite{methontology97} modified using another ontology design workflow introduced recently in \cite{POVEDAVILLALON2022104755}.\textit{**Copyright Esteban Guerrero.}}
    \label{fig:describe_methonto}
\end{figure}

Figure \ref{fig:describe_methonto} illustrates the phases of METHONTOLOGY adopted in this work to build the ontology. The ontology was developed using all the five stages. The integration stage is embedded within the first and second phases of specification and conceptualisation. 

\section{Formalization of Contradictions}\label{sec:formal_rep}

\subsection{Defining Object Properties and First Order Logic Formalism}\label{sec:defineObjProp}
\textit{Object} is an instantiation of desires and intentions of an entity. 
Objects exist in a hierarchy. At the highest complexity level the object is the \textit{motive} that an entity wants to attain. At lower levels its the \textit{goals} of actions, through which the subject is interacting with the world, which can also be represented as physical affordances (as defined by Gibson \cite{greeno1994gibson}) in this article \cite[Chp. 3]{kaptelinin2009acting}. 
Then an object can formally be defined as:

\begin{definition}[\textbf{Object}]\label{def:Obj}
   Let $D_n$ be the set of all the desires of an entity, $I_n$ be the set of all the intentions, $G_n$ represent goals of an entity, and $P_n$ be all the physical affordances (\cite{greeno1994gibson}). The object is composed of multiple levels, each one comprised of needs, desires, intentions, goals and physical affordances. Each level $n$ of the object can be described as a tuple $\langle \mathcal{N}_n, D_n, I_n, G_n, P_n \rangle$, where:\\
    $\mathcal{N}_n$ : the set of needs at level $n$ \\
    $D_n$ : the set of desires at level $n$ \\
    $I_n$ : the set of intentions at level $n$ \\
    $G_n$ : the set of goals at level $n$ \\
    $P_n$ : the set of affordances at level $n$

    Levels of abstraction of the object are connected from a level to the next, by the following function:\\
    $f_{N}: \mathcal{N}_n \rightarrow D_{n+1}$\\
    $f_{N}: D_n \rightarrow I_{n+1}$\\
    $f_{N}: I_n \rightarrow G_{n+1}$\\
    $f_{N}: G_n \rightarrow P_{n+1}$\\
    $\mathcal{N},\ D,\ I,\ G,\ P$ have their own sets
\end{definition}



\textit{Actors} are sentient entities (specifically, humans) living and acting in the world to fulfil their needs. An actor becomes a \textit{subject} when they have both \textit{``the ability and need to act"} transforming the world and the self by acting in it. Subject can be an individual or subgroup \cite{Engestrom_Sannino_CHAT_2018}. We expand the definition of \textit{subject} to include \textit{insentient} entities such as software, digital and embodied agents driven by artificial intelligence that have agency, and adopt or support  human goals. These insentient entities can transform the world, but may or may not have the capability to change themselves. 

\begin{definition}[\textbf{Subject}]\label{def:subject}
    Let actors $A$ be the set of all the sentient (humans and animals) $S = \{s_1,s_2\dots,s_i\}$ and insentient entities (software, virtual, embodied, and digital agents) $E = \{e_1,e_2\dots,e_j\}$, such that $A \subseteq S \cup E$.  Then a specific subject is an actor given as $sbj \in A$. 
    
\end{definition}

    \textit{Mediating Artefacts} are the symbols, signs, and understanding of the concepts used by the subject to \textit{mediate} with the objects \cite{BatiiBew2019}. They have a function to reverse the action, transforming the psychological operations to new forms, allowing the humans to \textit{``control their behaviour from the outside''} \cite[pp. 39-40]{vygotsky_1978}. 
    
    Mediating artefacts can be distinguished between technical and psychological artefacts. Both of these are mediators in the activity. 
    \\ \textbf{Technical Tools} are used to affect other things.  
    \\ \textbf{Psychological Tools} are signs used by the subject to affect other entities or themselves \cite[pp. 42]{kaptelinin2009acting}, \cite{Vygotsky1982}. Psychological tools can be physical artefacts such as art and diagrams or symbolic systems such as mathematical symbols and language. \\

\begin{definition}[\textbf{Mediating Artefacts}]\label{def:tools}
    Let $PT$ be the set of all psychological artefacts and $TE$ be the set of all the technical artefacts, then mediating artefacts $\mathcal{MT}$ is the union given as, $\mathcal{MT} \subseteq TE \ \cup \ PT$.
\end{definition}

    \textit{Activity system model (ASM)} is an \textit{``object-oriented, collective, and culturally mediated human activity, or activity system. \cite{Engestrom_Miettinen_Punamaki_1999}.''} Represents a model of collective activities, where the object is shared by a community and collectively carried out. The activity is an interaction between the subject, the object, and the community. These interacting components are mediated through mediating artefacts, rules, and division of labour \cite{Engestrom_Sannino_CHAT_2018}. A dialogue-based HRI is structured in this work as an ASM. 


    The \textit{community} is a group composed of the actors and other entities that have a shared object. For instance in a dialogue-based HRI scenario in healthcare, the robot, the human needing care, and other care providers are part of a community, sharing a common object to supporting the wellbeing of the human. 

    \textit{Rules} are the mediating components between the subject and the community. They are explicit and implicit rules, norms, and conventions regulating the activity \cite{kaptelinin2009acting}, such as in a medication scenario, medicines should be followed according to the prescription, dialogues between the human and the robot should follow the strategies set out by the designers.

    \textit{Division of labour} are the mediating component between the object and the community. The subject uses this component to coordinate their tasks with the community, creating a mediation between the community and the object (i.e., the motive) \cite{kaptelinin2009acting}. In a collaborative dialogue-based HRI healthcare scenario, a registered doctor prescribes medication, the human takes the medications, and the robot supports the human in this activity, through reminders and . 

    \textit{Outcome} is the intended result generated by the transformation of the object through the collaborative activity. The outcome in dialogue-based HRI healthcare scenario will be managing medications successfully. 

\begin{definition}[\textbf{Collaborative Activity}]\label{def:collabActvty}
    Let a dialogue-based HRI be an ASM, represented as a collaborative activity $ca$. Let the collaborative activity be an interaction between a community of formal and informal care providers $comm$, the subject (human or the robot) $sbj$ and the object (supporting health are wellbeing) $obj$, mediated by artefacts (medication box, prescription, dialogue strategies) $mt \in \mathcal{MT}$. The socio-cultural context $SOC$ consisting of the set of rules $R$ and the set of division of labour $DOL$ for a given activity, $SOC \subseteq R \ \cup \ DOL$. Then a collaborative activity at a given time point $tp$ is given as a function $ca(mt,\ sbj,\ o,\ comm,\ soc,\ tp) = \langle sbj', o' \rangle$.   
\end{definition}

\subsection{FoL formalism of ASM}
We use the object properties defined in Section\ref{sec:defineObjProp} and relationships from the foundational ontology we are developing, called Activity Theory-based Foundational Ontology (ATFOt) for FoL formalisation of ASM (Definition \ref{def:collabActvty}). Prefix \textit{atot} is used to refer to the foundational ontology in the rest of the formalisation. 

\begin{definition}
    \begin{center}
    $ DialHRI(ca) \equiv atot.M(ca)  
    \land \exists \ (x, y, obj, mt, soc, t(x \neq y), (x,y \in comm))  
    \land atot.Subject(x)  
    \land atot.hasNeeds(n,x) 
    \land atot.Subject(y) \land   
    \land atot.AdoptsNeeds(an,y) 
    \land atot.hasObject(obj)
    \land atot.hasTools(mt) 
    \land atot.hasContext(soc)
    \land atot.hasTimeInterval(t)
    \land atot.Employs(s,mt,obj) \land 
    \exists \ (z) (atot.Subject(z)
    \land atot.adoptsNeeds(n,z)) \rightarrow atot.adheresTo(z,soc) \land atfot.Transforms(obj,out) $     
    \end{center}
      
\end{definition}

The definition reads as follows: In an ASM, a subject \textit{(s)} either with intrinsic needs (sentient) \textit{(n)} or by adopting the needs of the others (insentient)\textit{(an)}, uses mediating artefacts \textit{(mt)} within a given socio-cultural context \textit{(soc)} on the object \textit{(obj)}. The subject and the community \textit{(z)} adhering to the the social context \textit{(soc)}, transform the object to achieve the desired outcome \textit{(out)}.

    \textit{Contradictions} \textit{``are historically emergent and systemic phenomena. Contradictions must therefore be approached through their manifestations. \cite[pp. 8]{Engestrom_Sannino_CHAT_2018}.'' } We define them as situations arising due to the discrepancies within the central ASM or with the members of the ASM family. 
    Contradictions are opportunities to learn and develop a better understanding of the activity and other components of an ASM \cite{tewari2023breakdown}. Internal contradictions lead to transformative change in the components that triggers the activity to change \cite{Schröder02082020}. Contradictions manifest in the form of a conflict, inconsistency, paradox, tension, or a dilemma \cite{Engestrom_Sannino_CHAT_2018}.

    \textit{Primary Contradictions} reside within the components of an ASM \cite{Engestrom_Sannino_CHAT_2018}, such as a missing artefact, or when it breaks down.

    \textit{Secondary Contradictions} emerge between different constitutive components of an ASM \cite{Engestrom_Sannino_CHAT_2018}, such as due to lack of knowledge about the rules or how to use the mediating artefact.

    \textit{Tertiary Contradictions} \textit{``appear when the activity system is reshaped and the new pattern collides with vestiges of the old one, generating resistance and forcing the new model to be modified'' \cite{Engestrom_Sannino_CHAT_2018}.} When new technologies are introduced in old processes, transforming the activity to a culturally more-advanced activity. Fear, excitement, and uncertainty when a robot is introduced in a care home for the first time.

    \textit{Quaternary Contradictions} \textit{``Takes shape when the transformed activity system interacts with its partner activities, generating tensions, disturbances and innovations in the network relations of involved activity systems'' \cite{Engestrom_Sannino_CHAT_2018}}. For example, in an HRI collaboration between a human and a robot, intended to support in healthcare management, a conflict between the human and the robot about following the doctor's prescription, could negatively impact human agency and autonomy. This is a contradiction with the existing rules around prioritizing human agency and autonomy in HRI.

\begin{definition}[\textbf{Contradiction}]\label{def:Contra}
    If $SM$ is the Engestrom's Activity System Model (ASM), then let a set of objects in the supporting ASMs be represented as $ (obj_p,\dots,o_q \in SM_{support})$. Let $(s_i,\ s_j \in SM_{central})$ be the two random components of the central ASM. Let $obj_n \in SM^{central}$ be the object of the central ASM, and let $obj_m \in SM_{advanced}$ be the object of the culturally more advanced ASM. Then we define four types of contradictions $Contra$ as follows:
    \begin{enumerate}
        \item [] \textbf{Primary Contradiction:} within a specific component of the central ASM $Contra_{pri}(s_i)$
        \item [] \textbf{Secondary Contradiction:} between a pair of components of central collaborative activity \\ $Contra_{sec}(s_i,s_j)$
        \item [] \textbf{Tertiary Contradiction:} is between the object of the central ASM and the culturally more advanced ASM $Contra_{quat}(obj_p,\\ obj_k)$.
        \item [] \textbf{Quaternary Contradiction:} is between the object of the central ASM and one of the supporting ASMs' $Contra_{ter}(obj_k, obj_m)$.
    \end{enumerate}
\end{definition}

A contradiction is an event that causes a \textit{focus shift} (fs) from the object of the central ASM, denoted as $o_i \in S$ to the alternative ASM with another object $o_k \in S$ where the contradiction will be resolved \cite{TewaLindFrontiers2022}. As such, contradictions can be represented as a mapping function:

\begin{equation}
    fs: S \rightarrow S
\end{equation}

\subsubsection{FoL Formalism of Contradictions}
The formalism of contradictions is based on a formal specification of three novel principles that rule every collaborative human activity. These principles are relevant for the computational reasoning about an ongoing activity represented in an ontology. Informally speaking these guiding principles define some of the transformations in the ontology due to contradictions, as follows:

\begin{itemize}
	   \item[] Principle 1: If there is a contradiction in the central ASM, it transforms to be the object, transitioning the focus to an alternative ASM.
       
       \begin{axiom}{A1:}
       \begin{center}
        $\forall (c) \ atot.Contra(c) \land 
        \exists (obj) \ atot.SM_{central}(obj) 
        \rightarrow atot.SM_{alternative}(obj') \land \neg atot.SM_{central}(obj).$   
       \end{center}
       \end{axiom}
	   \item[] Principle 2: If there is a contradiction in the central ASM, then there is a conflict with the ASM family.
       \begin{axiom}{A2:}
           \begin{center}
               $\forall (c) \ atot.Contra(c) \land \exists (ca) \ atot.SM_{central}(ca) \rightarrow \exists (ca) \ atot.SM_{family}(ca) 
               \land atot.Conflict(SM_{family}(ca)) \land \neg SM_{central}(ca).$ 
           \end{center}
       \end{axiom}
	   \item[] Principle 3: If there is a contradiction in the central ASM then there is a contradiction in its inherited ASMs, i.e., the culturally more-advanced ASM and its related ASM family.
       \begin{axiom}{A3:}
           \begin{center}
               $\forall (c) \ atot.Contra(c) \land \exists (ca) \ atot.SM_{central}(ca) \rightarrow \exists(c) \ Contra(c) \land \exists(ca) SM_{inherit}(ca)$
           \end{center}
       \end{axiom}
\end{itemize}

\section{Conclusion and Ongoing Work}\label{sec:future_work}
This research presented preliminary results on the ontological formalism of dialogue-based interactions for HRI and HAI and related contradictions. Engestrom's Activity System Model (ASM) and the concept of contradictions from Activity Theory were used as a foundational theory in the conceptualisation and definition of contradictions and dialogue-based interactions. 

\textit{This research contributed with a novel formalisation of contradictions concepts and three novel principles guiding the transformations in an ontology caused by the inherent contradictions.  
}
The presented work is part of an ongoing research on developing a foundational ontology called \textit{activity theory-based foundational ontology (ATFOt)}. ATFOt represents collective activity and related contradictions such as errors, failures, conflicts, and knowledge issues in dialogue-based interactions between humans and agents. 

Currently, we are finishing the formalisation and implementation of ATFOt using set theory, FoL, and Protege environment, respectively. The implemented ontology will be evaluated qualitatively in studies and verified using coverage and competency questions. 

ATFOt can be used by agents intended to interact with humans to formally represent, reason about, and manage dialogues and related contradictions. This ontology is a step towards reliable, transparent, and trustworthy dialogue management strategies for both learning and rule-based agents (embodied, digital, or virtual.) This ontology can be used to standardise the terminology for dialogues and relevant contradictions. This standardisation advances the research in contradiction management by enabling the reusability and interoperability of the terminologies across domains in human-agent dialogues.  

\section{Funding}
Funding was received for conducting research and authorship of this article. The research was supported by HumanE-AI-Net project, which received funds from the European Union's Horizon 2020 research and innovation programme (grant agreement 952026) and by the European Union under Horizon EU project LearnData, 101086712.

\section{Acknowledgment}
The authors are grateful to Helena Lindgren, Esteban Guerrero, and Vera C. Kaelin for their invaluable contribution and insights to identifying the research gap, conceptualization of the idea, in selecting the theoretical framework and the methodology, and in initial theoretical formulations.

\bibliographystyle{IEEEtran}
\bibliography{JWOSMARS_WrkShp_Aprl2026}

\end{document}